\documentclass[a4paper,11pt]{article}
\pdfoutput=1 

\usepackage{jheppub} 
\usepackage{amsmath,amsthm,amssymb,amsfonts}

\usepackage{caption}
\usepackage[utf8]{inputenc}
\usepackage{graphicx}
\usepackage{etoolbox}

\title{\boldmath Virtual Exchange of Continuous Spin Fields 
}

\author{Victor O. Rivelles}
\affiliation{ Instituto de F\'{i}sica, Universidade de S\~ao Paulo,\\ Rua do Mat\~ao, 1371, 05508-090  S\~ao Paulo, SP, Brazil}

\emailAdd{rivelles@fma.if.usp.br}

\abstract{We discuss the long range interactions mediated by continuous spin fields. We start by deriving the propagator for a continuous spin field using the antifield BRST formalism. Then we couple the continuous spin field to a conserved current to find the interaction energy due to static disturbances of the vacuum. For sources having charges of the same sign we find an attractive force at small distances and a repulsive force at large distances. We also discuss the conserved current degrees of freedom.

}

\begin{document} 
\maketitle
\flushbottom

\section{Introduction \label{s0}}

All known elementary particles are classified according to the unitary irreducible representations of the Poincaré group and the interactions among them are successfully described by quantum field theories in Minkowski spacetime. However there is a gap which needs to be fulfilled. The continuous spin particle is a massless unitary irreducible representation of the Poincaré group which is characterized by a non-vanishing quartic Casimir operator having eigenvalue $\rho^2$ with $\rho$ having mass dimension one \cite{Wigner:1939cj}. In the context of quantum field theory it presents serious troubles \cite{Yngvason:1970fy,Iverson:1971hq,Chakrabarti:1971rz,Abbott:1976bb,Hirata:1977ss}. Only recently a classical field theory for continuous spin particles was proposed not in spacetime but on a cotangent bundle over Minkowski spacetime where the field depends not only on the spacetime coordinates $x^\mu$ but also on an extra coordinate $\eta^\mu$ \cite{Schuster:2014hca}. 
Like the other massless particles the continuous spin field has a gauge symmetry but now on the cotangent bundle not in spacetime. 
The fact that we have to work on a cotangent bundle brings new issues which are absent in conventional quantum field theory.  In this work we present the first steps towards a systematic quantization of bosonic continuous spin fields by deriving the static energy generated by the virtual exchange of continuous spin fields.

In order to do that we must make sure that continuous spin particles can interact with other particles. As argued in \cite{Rivelles:2016rwo} we can couple an abelian gauge field to a complex continuous spin field preserving the abelian gauge symmetry. However the local symmetry of the continuous spin field is completely lost. This was expected because when the value of continuous spin goes to zero the continuous spin field action reduces to a sum of actions for all integer values of the spin which, of course, do not propagate in flat spacetime. If we restrict to cubic couplings involving one continuous spin field and  two massive scalar fields \cite{Bekaert:2017xin,Rivelles:2018tpt} or one continuous spin tachyon and two massive scalar fields \cite{Rivelles:2018tpt} then conserved currents can be found on the cotangent bundle. More general cubic vertices involving massless or massive fields of any spin and continuous spin fields or continuous spin tachyons also have been built in a light-cone gauge formulation \cite{Metsaev:2017cuz,Metsaev:2018moa}. More recently a worldline coupling of spinless matter particle to continuous spin fields was also analyzed \cite{Schuster:2023xqa}.

The first step to quantize the continuous spin field is to find its propagator. Adding a source term to the action and removing pure gauge terms from the equations of motion provides a quick way to find the propagator \cite{Bekaert:2017xin}. But since the action for the continuous spin field  has a gauge symmetry which is reducible in the sense that its gauge parameters also have a gauge symmetry means that we must be more careful. One way of doing this is by using the BRST approach. Several approaches of BRST symmetry were developed and applied to study properties of the continuous spin case. Treating the Wigner equations for continuous spin particles as a constrained system allowed a formulation of the BRST symmetry developed in \cite{Bengtsson:2013vra,Buchbinder:2018yoo,Buchbinder:2020nxn,Buchbinder:2022msd}. Considering the continuous spin field by itself as a constrained system leads to the BRST formulation developed in \cite{Alkalaev:2017hvj,Alkalaev:2018bqe} and to the BRST-BFV approach of \cite{Burdik:2019tzg,Metsaev:2018lth}. In this paper we will take into account the fact that the classical gauge theory is reducible and apply the antifield BRST formalism\footnote{For a review of the antifield BRST formalism and original references see \cite{Fuster:2005eg,Gomis:1994he,Henneaux:1989jq}} to the original classical action defined on a cotangent bundle \cite{Schuster:2014hca} leading to a gauge fixed action with a manifest nilpotent BRST symmetry. We then find that for a simple gauge choice the propagator goes like $1/k^2$ as for other massless particles and also that it does not depend on $\rho$. 

We then set up the path integral formalism in the cotangent bundle over Minkowski spacetime to get the generating functional for connected diagrams. To this end we add external sources minimally coupled to the continuous spin field. They will be used to find out the interaction energy due to localized static disturbances of the vacuum. To do that we need to consider a conserved current 
in the cotangent bundle. We find that the current is not point like as in conventional gauge theories but it is spread around a point in space and has a constant flux on the cotangent bundle. We then find the classical potential energy between two charged particles. Charges of the same sign go like $-1/r^{D-3}$ at small distances and it is independent of $\rho$. At large distances we find $\alpha^{(D-2)/2} \exp(-\alpha r)/r^{(D-4)/2}$ with $\alpha$ proportional to $\rho$.  
So at small distances particles with charges of the same sign attract each other while at large distances they repel each other. Quite interestingly this has a flavor of the dark side of the universe where at small distances we find attraction and at large distances we find repulsion of ordinary matter.

Next we consider the exchange of continuous spin fields between the two sources. Firstly we solve the current conservation equation to identify the independent components of the current when it is expanded in $\eta^\mu$ finding that the components are traceless in their transversal $SO(D-2)$ indices as expected. However when we compute the generating functional for connected diagrams we get the product of three summations involving the components of the two currents instead of a single summation which is expected if each helicity appears just once. However, taking the first few terms of the three summations we find that they remarkably simplifies to a single summation only. We then construct a new generating functional with a single summation of the product of the two traceless $SO(D-2)$ currents leaving the coefficients of the summation arbitrary. Comparing  the first few terms of each functional suggests a general pattern for the coefficients of the new functional. Unfortunately it is not possible to compare both functionals analytically but only in a case by case situation. In all tested cases we found the same result for both functionals giving a very strong support that they are really the same. 

The contents of the paper is as follows. In Section 2 we get the propagator for the continuous spin field using the antifield BRST formalism. Then we obtain the continuous spin propagator in Section 3 and derive the potential energy due to static forces generated by the exchange of virtual continuous spin fields in Section 4. In Section 5 we discuss the degrees of freedom of the conserved current and in  Section 6 we present  some final comments.

\section{Antifield BRST Formalism for Continuous Spin Fields \label{s1}}

The field for a continuous spin particle is defined on a $D$-dimensional cotangent bundle with coordinates $x^\mu$ and $\eta^\mu$, $\mu=0, \dots, D-1$ and its action is given by \cite{Schuster:2013vpr} 
\begin{equation}\label{2.1}
	S = \frac{1}{2} \int d x \, d\eta \, \delta^\prime(\eta^2+\mu^2) \left( (\partial_x \Psi(\eta,x))^2 - \frac{1}{2} (\eta^2+\mu^2) \left( \Delta \Psi(\eta,x) \right)^2 \right), 
\end{equation}
where $\Delta = \partial_\eta \cdot \partial_x + \rho$, 
$\rho$ is the real continuous spin parameter 
and $\delta^\prime$ is the derivative of the delta function with respect to its argument. The metric is mostly minus. Integrations over $\eta^\mu$ are divergent so they must be regulated. As in \cite{Schuster:2014hca} we use a Wick rotation to perform all $\eta^\mu$ integrals. The factor $\mu^2$ is introduced to track dimensions of $\eta^\mu$ and to make dimensional analysis easier. It can always be set equal to 1 by rescaling  $\eta^\mu$. Notice also that $\mu\rho$ has dimension of inverse length and it is the only parameter in the free theory. The derivative of the delta function constrains the dynamics to the hyperboloid $\eta^2+\mu^2=0$ and its first neighborhood. 

The action is invariant under the following global transformations: spacetime translations, Lorentz transformations and a $\eta^\mu$ dependent translation along $x^\mu$ given by $\delta x^\mu = \omega^{\mu\nu}\eta_\nu$, with $\omega^{\mu\nu}$ antisymmetric. This last symmetry does not preserve the natural symplectic structure of the cotangent bundle \cite{Rivelles:2016rwo}. The action is also invariant under the following local transformations
\begin{equation}\label{bbb5}
	\delta \Psi(\eta,x) = \left( \eta\cdot\partial_x - \frac{1}{2} (\eta^2+\mu^2) \Delta \right) \epsilon(\eta,x)  + \frac{1}{4} (\eta^2+\mu^2)^2 \chi(\eta,x),
\end{equation}
with $\epsilon(\eta,x)$ and $\chi(\eta,x)$ being the local parameters. Fields defined in the cotangent bundle can be expanded around the hyperboloid and the role of the $\chi$ symmetry is to remove all components of such an expansion except for the first two so that it restricts the propagation of $\Psi$ to the $\eta^2+\mu^2$ hyperboloid and its first neighborhood. On the other side, the $\epsilon$ symmetry is an usual gauge symmetry removing gauge degrees of freedom. These local symmetries are reducible \cite{Rivelles:2014fsa} since 
\begin{eqnarray}
	\delta \epsilon(\eta,x) &=& \frac{1}{2} (\eta^2+\mu^2) \Lambda(\eta,x), \label{b1} \\ 
	\delta \chi(\eta,x) &=& \Delta \Lambda(\eta,x), 	\label{b2}
\end{eqnarray}
leave (\ref{bbb5}) invariant. This symmetry mimics the $\chi$ symmetry for $\Psi$ and can be used to limit the expansion of $\epsilon$ around the hyperboloid $\eta^2 + \mu^2 =0$ to just the first term of the expansion. More details can be found in \cite{Rivelles:2018tpt}.

Since we have local symmetries which are reducible we have to introduce ghost fields in such a way that the reducibility is taken into account. We will use the BRST field-antifield formalism  where the original local symmetries are replaced by a rigid nilpotent BRST symmetry \cite{Fuster:2005eg,Gomis:1994he,Henneaux:1989jq}. This is achieved in two steps. Firstly we introduce ghost fields and antifields on the cotangent bundle in such a way that a BRST symmetry is manifest and the action is still invariant under the local transformations (\ref{bbb5}). Then we introduce a gauge fixing fermion which will require  additional fields and antifields. These additional fields and antifields are added keeping the nilpotency of the BRST transformation. Once this is done we can make a choice for the gauge fixing fermion to obtain the propagator.

We start by introducing ghost fields for each local symmetry of $\Psi(\eta,x)$. We need two fermionic ghost fields $c_\epsilon(\eta,x)$ and $c_\chi(\eta,x)$ both with ghost number 1, and one additional bosonic ghost field $c_\Lambda(\eta,x)$ with ghost number 2, all of them required by the reducible structure (\ref{b1}) and (\ref{b2}). Then we introduce the corresponding antifields, a fermionic antifield $\Psi^*(\eta,x)$ with ghost number -1, two bosonic antifields $c^*_\epsilon(\eta,x)$ and $c^*_\chi(\eta,x)$ both with ghost number -2, and a fermionic antifield $c^*_\Lambda(\eta,x)$ with ghost number -3. They are listed in Table 1 and Table 2.

\vspace{.75cm}

\begin{minipage}[c]{0.4\textwidth}
\centering
\begin{tabular}{ l | c | c  }
                       & \text{statistics} & \text{ghost number} \\
\hline
 $\Psi(\eta,x)$         & \text{bosonic}    & 0 \\
 $c_\epsilon(\eta,x)$  & \text{fermionic}  & 1 \\
 $c_\chi(\eta,x)$      & \text{fermionic} & 1 \\
 $c_\Lambda(\eta,x)$   & \text{bosonic}    & 2 \\
\end{tabular}
\captionof{table}{Fields 
}
\end{minipage}
\begin{minipage}[c]{0.5\textwidth}
\centering
\begin{tabular}{ l | c | c  }
                     & \text{statistics} & \text{ghost number}\\
\hline
$\Psi^*(\eta,x)$      & \text{fermionic}  & -1 \\
$c^*_\epsilon(\eta,x)$ & \text{bosonic}    & -2 \\
$c^*_\chi(\eta,x)$     & \text{bosonic}    & -2 \\
$c^*_\Lambda(\eta,x)$  & \text{fermionic}  & -3  \\
\end{tabular}
\captionof{table}{Antifields}
\end{minipage}

\vspace{.75cm}

Then we can build  the action
\begin{align}\label{b3}
	S_1 = \int d x \, d\eta \left\{ \frac{1}{2} \delta^\prime(\eta^2+\mu^2)   \left( (\partial_x \Psi(\eta,x))^2 - \frac{1}{2} (\eta^2+\mu^2) \left( \Delta \Psi(\eta,x) \right)^2 \right) 
	\right. \nonumber \\
	\left. + \Psi^* \left[ \left( \eta\cdot\partial_x - \frac{1}{2}(\eta^2+\mu^2)\Delta \right) c_\epsilon + \frac{1}{4}(\eta^2+\mu^2)^2 c_\chi \right] + \left( 
	c^*_\chi \Delta + \frac{1}{2}(\eta^2+\mu^2) c^*_\epsilon \right) c_\Lambda  \right\},
\end{align}
which is invariant under the nilpotent BRST transformations\footnote{Our convention for the BRST transformation for the product of two fields is $\delta(XY) = X \delta Y + (-1)^{\epsilon _Y} (\delta X ) Y$, where $\epsilon_Y$ is the grading of $Y$.}

\begin{align}
	&\delta \Psi =  \left( \eta\cdot\partial_x - \frac{1}{2}(\eta^2+\mu^2)\Delta \right)c_\epsilon + \frac{1}{4}(\eta^2+\mu^2)^2 c_\chi, \label{b4} 
	\\
	&\delta c_\epsilon = \frac{1}{2}(\eta^2+\mu^2) c_\Lambda, \label{b5} \\
	&\delta c_\chi = \Delta c_\Lambda, \label{b6}\\
	&\delta\Psi^* = \delta^\prime(\eta^2+\mu^2) \left( \Box_x - \eta\cdot\partial_x\Delta + \frac{1}{2}(\eta^2+\mu^2)\Delta^2 \right) \Psi, \label{b7} \\
	&\delta c^*_\epsilon = -2 \left( \eta\cdot\partial_x + \frac{1}{4}(\eta^2+\mu^2)\Delta \right) \Psi^*, \label{b8} \\
	&\delta c^*_\chi  = \frac{1}{4}(\eta^2+\mu^2)^2 \Psi^*, \label{b9} \\
	&\delta c^*_\Lambda = - \Delta c^*_\chi - \frac{1}{2}(\eta^2+\mu^2) c^*_\epsilon. \label{b10}
\end{align}
Notice that the terms in the second line of (\ref{b3}) are not restricted to propagate on the hyperboloid and its first neighbourhood. On the other side, 
the BRST transformation of $\Psi^*$ (\ref{b7}) is restricted to the hyperboloid and its first neighbourhood. This structure is crucial, for instance, to show that the BRST transformation of $c^*_\epsilon$ is indeed nilpotent. Notice also that (\ref{b3}) is still invariant under the local transformations (\ref{bbb5}) so that no gauge fixing was done yet.

The next step is to gauge fix (\ref{b3}). In order to do that we need the introduction of further fields and antifields as follows. Due to the $\epsilon$ symmetry we add the pair $\overline{c}_\epsilon(\eta,x)$ and $b_{\epsilon}(\eta,x)$, the first a fermionic and the second a bosonic field, with ghost number -1 and zero, respectively. In a similar way, for the $\chi$ symmetry we add $\overline{c}_\chi(\eta,x)$ and $b_{\chi}(\eta,x)$, fermionic and bosonic fields respectively, also with ghost number -1 and zero.  We also need a pair $\overline{c}_\Lambda(\eta,x)$ and $b_\Lambda(\eta,x)$, bosonic and fermionic fields respectively, with ghost number -2, and -1. Finally the last pair $\sigma(\eta,x)$ and $\pi(\eta,x)$, bosonic and fermionic fields respectively, have ghost number zero and 1. Then we have to add the corresponding antifields $\overline{c}^*_\epsilon, b^*_\epsilon, \overline{c}_\chi, b^*_\chi, \overline{c}^*_\Lambda, b^*_\Lambda, \sigma^*$ and $\pi^*$ with the opposite statistics to the former ones and ghost number 0, -1, 0, -1, 1, 0 -1 and -2, respectively. They are all listed in Table 3 and Table 4. 

\vspace{.5cm}

\begin{minipage}[c]{0.5\textwidth}
\centering
\begin{tabular}{ l | c | c  }
                       & \text{statistics} & \text{ghost number} \\
\hline
 $\overline{c}_\epsilon(\eta,x)$  & \text{fermionic}  & -1 \\
 $         {b}_\epsilon(\eta,x)$  & \text{bosonic}    & 0  \\
 $\overline{c}_\chi(\eta,x)$      & \text{fermionic}  & -1  \\
 $         {b}_\chi(\eta,x)$      & \text{bosonic}    & 0  \\
 $\overline{c}_\Lambda(\eta,x)$   & \text{bosonic}    &-2 \\
 $         {b}_\Lambda(\eta,x)$   & \text{fermionic}  & -1  \\
 $\sigma(\eta,x)$                 & \text{bosonic}    & 0  \\
 $\pi(\eta.x)$                    & \text{fermionic}  & 1  \\                 
\end{tabular}
\captionof{table}{Fields 
}
\end{minipage}
\begin{minipage}[c]{0.5\textwidth}
\centering
\begin{tabular}{ l | c | c  }
                     & \text{statistics} & \text{ghost number}\\
\hline
 $\overline{c}^*_\epsilon(\eta,x)$  & \text{bosonic}  & 0 \\
 $         {b}^*_\epsilon(\eta,x)$  & \text{fermionic}& -1 \\
 $\overline{c}^*_\chi(\eta,x)$      & \text{bosonic}  & 0  \\
 $         {b}^*_\chi(\eta,x)$      & \text{fermionic}& -1  \\
 $\overline{c}^*_\Lambda(\eta,x)$   & \text{fermionic}& 1 \\
 $         {b}^*_\Lambda(\eta,x)$   & \text{bosonic}  & 0  \\
 $\sigma^*(\eta,x)$                 & \text{fermionic}& -1 \\
 $\pi^*(\eta.x)$                    & \text{bosonic}  & -2 \\
\end{tabular}
\captionof{table}{Antifields}
\end{minipage}

\vspace{.5cm}

\noindent 
They are used to write the action  
\begin{align}\label{b11} 
	S_2 = \int dx \, d\eta \left( \overline{c}^*_\epsilon b_\epsilon + \overline{c}^*_\chi b_\chi + \overline{c}^*_\Lambda b_\Lambda + \sigma^* \pi \right), {\color{red} 
	}
\end{align}
which is trivially invariant under the BRST transformations
\begin{align}
	\delta\overline{c}_\epsilon &= b_\epsilon, \qquad\qquad \delta\overline{c}_\chi = b_\chi, \label{b12}  
	\\
	\delta\overline{c}_\Lambda &= b_\Lambda, \qquad\qquad  \delta\sigma = \pi, \label{b13} \\
	\delta b^*_\epsilon &= - \overline{c}^*_\epsilon, \qquad\quad\, \delta b^*_\chi = - \overline{c}^*_\chi, \label{b14} \\
	\delta b^*_\Lambda &= \overline{c}^*_\Lambda, \qquad\qquad \delta \pi^* = \sigma^*, \label{b15}
\end{align}
so that  $S_1+S_2$ is also BRST invariant. The action $S_2$  has no dependence on $\delta^\prime$.

Now we introduce the gauge fixing fermion ${\bf\Omega}$ which is a quadratic function of fields, antifields and their derivatives, has ghost number -1 and is fermionic. It will be used to eliminate  all the antifields by imposing the condition
\begin{align}\label{b16}
	\Phi^*_A = \frac{\partial \bf\Omega}{\partial \Phi_A}, 
\end{align}
where $\Phi_A$ is the set of all fields and $\Phi^*_A$ the set of all antifields. A suitable choice for the  gauge fixing fermion depends on two real parameters $\zeta$ and $\xi$ and is given by 
\begin{align} \label{b17}
	{\bf\Omega} = \int dx \, d\eta \,  \delta(\eta^2+\mu^2) \left[ \overline{c}_\epsilon (\Delta \Psi + \sigma) + \overline{c}_\chi \left( \Psi + \frac{1}{\zeta} b_\chi + \frac{1}{\sqrt{\zeta\xi}} \sigma \right) + b_\Lambda  \sigma + \overline{c}_\Lambda c_\chi \right]. 
\end{align}
Using (\ref{b16}) we can eliminate all antifields so that the $S_1+S_2$ reduces to the gauge fixed action
\begin{align}\label{b18}
S_{gf} = \int dx \, d\eta \,  \delta^\prime(\eta^2+\mu^2) \times  \hspace*{9.5cm} \nonumber \\
\times \left\{  \frac{1}{2}(\partial_x \Psi(\eta,x))^2 
 -	(\eta^2+\mu^2) \left[ \frac{1}{4} (\Delta \Psi)^2 + (\Delta \Psi + \sigma) b_\epsilon + (\Psi + \frac{1}{\zeta} b_\chi + \frac{1}{\sqrt{\zeta\xi}}\sigma ) b_\chi   \right.\right. \hspace*{0.5cm}  \nonumber \\
\left.\left. +  ( c_\chi - \pi ) b_\Lambda + \overline{c}_\epsilon ( \Box_x c_\epsilon + \pi) +  \overline{c}_\chi ( \eta\cdot\partial_x c_\epsilon + \frac{1}{\sqrt{\zeta\xi}} \pi ) + \overline{c}_\Lambda \Delta c_\Lambda  \right] \right\}. 
\end{align}
Notice that the two first terms are those of the classical action (\ref{2.1}) while the remaining terms involve all ghost fields. The gauge fixed action (\ref{b18}) is invariant under the nilpotent BRST transformations (\ref{b4}), (\ref{b5}), (\ref{b6}), (\ref{b12}) and  (\ref{b13}).

\section{Continuous Spin Field Propagator \label{2a}}

In order to find the propagator we have to consider the path integral formulation on a cotangent bundle. So let us consider a path integral involving a field $\Xi(\eta,x)$ on a cotangent bundle, an operator $\cal{O}$ involving spacetime derivatives, $\eta^\mu$ and its derivatives, and  also a  delta function $\delta(\eta^2+\mu^2)$. We then get  
\begin{align}\label{h1}
	\int\, {\cal D}\Xi \,\,  e^{i \int dx\, d\eta \, \delta(\eta^2 + \mu^2) \, \Xi (\eta,x) \, {\cal O} \, \Xi (\eta,x) } = det^{\mp 1/2} \left( \delta(\eta^2+\mu^2) \cal{O} \right), 
\end{align}
where the minus (plus) sign is for a bosonic (fermionic) field $\Xi(\eta,x)$. 
Using the gauge fixed action (\ref{b18}) we find that the integration over $c_\Lambda$ and $\overline{c}_\Lambda$ gives a factor of $det^{-1}\left( \delta(\eta^2+\mu^2) \, \Delta\right)$ in the path integral measure. Integration over $\pi$ and $\eta$ gives rise to two delta functions which can be used to solve for $b_\epsilon$ and $b_\Lambda$. Then integration over $b_\chi$ gives a factor of $det^{-1/2}\left( \frac{2}{\zeta} \delta(\eta^2+\mu^2) \right)$ and a contribution to the action of the form  $\frac{\zeta}{2} \left( (1 - \frac{\Delta}{\sqrt{\zeta}} ) \Psi \right)^2$. Finally integration over $\overline{c}_\epsilon$ and  $\overline{c}_\chi$ gives rise to delta functions which can be solved for $c_\chi$ and $c_\epsilon$  contributing with a factor of $ det\left( \delta(\eta^2+\mu^2) ( \eta\cdot \partial_x - \frac{\Box}{\sqrt{\zeta}} )\right)$. All together gives a factor of 
\begin{align}\label{h2}
det^{-1} \left( \delta(\eta^2+\mu^2)\Delta \right) \,\, det^{-1/2} \left(\frac{2}{\zeta} \delta(\eta^2+\mu^2) \right) \,\, det \left( \delta(\eta^2+\mu^2) \left( \eta\cdot\partial_x - \frac{\Box_x}{\sqrt\zeta} \right)\right),  
\end{align}
in the path integral measure
and the gauge fixed action (\ref{b18}) is reduced to 
\begin{align}\label{b19}
	   S_{gf} = \int dx \, d\eta \,\, \frac{1}{2} \delta^\prime(\eta^2+\mu^2) \times  \hspace*{7.5cm} \nonumber \\ 
\times \left\{  (\partial_x \Psi(\eta,x))^2	-	\frac{1}{2} (\eta^2+\mu^2) \left( (1 - \frac{1}{\xi}) (\Delta \Psi)^2 + 2 \sqrt{\frac{\zeta}{\xi}} \Psi\Delta\Psi -  \zeta \Psi^2 \right)\right\}.
\end{align}
This gauge fixed action is no longer invariant under the $\epsilon$ gauge transformation in (\ref{bbb5}) but it is still invariant under the $\chi$ transformation as expected. 

The propagator can be found by adding a source term ${J}(\eta,x)$ to the action (\ref{b19}) 
\begin{align}\label{b20}
	S_{J} = \int d\eta \, dx \, \delta^\prime(\eta^2+\mu^2) \, {J}(\eta,x) \, \Psi(\eta,x). 
\end{align}
The presence of the $\delta^\prime(\eta^2+\mu^2)$ makes $S_J$  invariant under 
\begin{align}\label{e2}
	\delta J(\eta,x) = \frac{1}{4} (\eta^2+\mu^2)^2 \,\, \Xi(\eta,x),
\end{align}
so that $J(\eta,x)$ is restricted to the hyperboloid $\eta^2 + \mu^2 =0$ and its first neighborhood. 
We can then find the field equation from (\ref{b19}) and (\ref{b20}), multiply it by $\eta^2 + \mu^2$ and apply $\Delta$ to get the generalization of the current conservation equation in a cotangent space \cite{Rivelles:2016rwo}
\begin{align}\label{e3}
	\delta(\eta^2+\mu^2) \Delta J(\eta,x) = 0.
\end{align}

The gauge choice which gives the simplest propagator is $\zeta=0, \xi=1$ 
so that the field equation in momentum space becomes 
\begin{equation}\label{b21}
	 \delta^\prime(\eta^2+\mu^2)  k^2 \tilde{\Psi}(\eta,k) = \delta^\prime(\eta^2+\mu^2)   \tilde{J}(\eta,k).  
\end{equation}
The propagator in momentum space, $\tilde{D}(\eta,k)$, is taken as $\tilde{\Psi}(\eta,k) = \tilde{D}(\eta,k) \, \tilde{J}(\eta,k)$ so that  the momentum space propagator is simply 
\begin{align}\label{b22}
\tilde{D}(\eta,k) = \frac{1}{k^2} + \frac{1}{4} (\eta^2+\mu^2)^2 \tilde{\varepsilon}(\eta,k),  
\end{align}
with $\tilde{\varepsilon}(\eta,k)$ arbitrary. The propagator then lives only on the $\eta^2+\mu^2=0$ hyperboloid and its first neighborhood but with no contribution to the first order neighborhood. This is expected since in the classical theory $\Psi(\eta,x)$ also does not propagate beyond the first neighborhood of the $\eta^2+\mu^2=0$ hyperboloid.

\section{
Virtual Exchange of Continuous Spin Fields \label{s3}}

As remarked before the path integral formalism for continuous spin fields has to be formulated on a cotangent bundle in such a way that the dynamics is constrained to the hyperboloid $\eta^2+\mu^2=0$ and its first neighborhood. 
Apart from that the procedure is similar to the case of ordinary fields in Minkowski spacetime. Then the generating functional for a classical external source $J(\eta,x)$ is taken to be 
\begin{align}\label{c1}
	Z[J] = \int{\it D} \Psi \,\, e^{i S_{gf}[\Psi] + i \int{dx \,d\eta \,\, \delta^\prime(\eta^2+\mu^2) \, J(\eta,x) \Psi(\eta,x)}} = e^{i \it{W[J]}}, 
\end{align}
where $S_{gf}$ is the gauge fixed classical action (\ref{b19}) 
and $W[J]$ is the generating functional for connected diagrams which can be written in momentum space as
 \begin{align}\label{c2}
	 W[J] = - \frac{1}{2} \int dk \, d\eta \,\, \delta^\prime(\eta^2+\mu^2) \, \tilde{J}(\eta,k) \tilde{D}(\eta,k) \tilde{J}(\eta, -k), 
 \end{align}
with $\tilde{D}(\eta,k)$ being the propagator for the continuous spin field in the gauge $\zeta=0, \xi=1$ (\ref{b22}). 

In general the static force among charged particles can be derived by the analysis of virtual particles exchanged by its sources\footnote{See \cite{Zee:2003mt} for instance}. In the present case continuous spin fields are the gauge fields carrying a force that intermediate  interactions among matter particles charged under this force. Then we can use the path integral formulation for continuous spin fields found in the previous section to find the classical force that acts on the matter particles. This can be achieved by computing the virtual particle exchange of continuous spin fields. In order to do so we take the source $J(\eta,x)$ as a matter current that interacts with the continuous spin field. 

So we assume that a virtual continuous spin field is created by a disturbance of the vacuum and is then absorbed by another disturbance of vacuum. These disturbances associated to the two charged particles are given by currents which are assumed to be static and localized in space at $\vec{x}_1$ and $\vec{x}_2$ 
\begin{align}\label{d3}
	J(\eta,\vec{x}) = J_1(\eta,\vec{x}-\vec{x}_1) + J_2(\eta,\vec{x}-\vec{x}_2), 
\end{align}
and must satisfy the conservation equation (\ref{e3}). We will also neglect current self-interactions so that (\ref{c2}) becomes
\begin{align}\label{c4}
	 W_{12}[J] = - \frac{1}{2} \int dk \, d\eta \,\, \delta^\prime(\eta^2+\mu^2) \left( \tilde{J}_1(\eta,k) \tilde{D}(\eta,k) \tilde{J}_2(-\eta, k) + \tilde{J}_2(\eta,k) \tilde{D}(\eta,k) \tilde{J}_1(-\eta, k)\right). 
 \end{align}
In this case $Z[J]$ gives the probability amplitude for the creation, propagation and annihilation  of the virtual continuous spin field which can be written in terms of its Hamiltonian operator ${H}$ as $<0| e^{-i H T}|0> = e^{-i E T}$, where $E$ is the energy change caused by the disturbance and $T$ the elapsed time. Then the interaction energy due to the disturbances of the vacuum is given by $E = - W_{12}/T$.

In order to compute the interaction energy due to static disturbances of the vacuum (\ref{c4}) we have to specify the currents $J_a(\eta,x), \, a=1,2$. The currents must be static and assumed to be concentrated at two different regions around $\vec{x}_1$ and $\vec{x}_2$. Since the total current $J(\eta,x)$ must be conserved (\ref{e3}) we assume that each current $J_a(\eta,\vec{x}_a)$ is also conserved. For spins 0 and 1 the current is simply proportional to $\delta(\vec{x}-\vec{x}_a)$ but 
in the continuous spin case this is non longer true 
and requires the addition of terms such that the current is not strictly localized at $\vec{x}_a$ but spread around $\vec{x}_a$. 
Besides that it should lead to a non vanishing exchange current (\ref{c4}). The current dependence  on $\eta$ is restricted by (\ref{e2}) so that it goes up to first order in $\eta^2  + \mu^2$. 
All this leads to a conserved current satisfying (\ref{e3}) given by 
\begin{align}\label{g1}
	&J_a(\eta,\vec{x}) =  \frac{2}{D-2} \, q_a \, \biggl[ \delta(\vec{x}-\vec{x}_a) -   \biggl.  
	\nonumber \\
&\left.	\frac{1}{(2\pi)^\frac{D-1}{2}} 
	\left( \rho \vec{\eta} \cdot\vec{\partial}_x + \frac{(\mu\rho)^2}{D-2} - \frac{1}{2} \rho^2(\eta^2+\mu^2) \right) \left( \frac{\mu\rho}{\sqrt{D-2}}  \frac{1}{ |\vec{x}-\vec{x}_a|} \right)^\frac{D-3}{2}  K_\frac{D-3}{2} \left( \frac{\mu\rho}{\sqrt{D-2}} |\vec{x}-\vec{x}_a| \right) \right],
\end{align}
where $K_i(x)$ is the modified Bessel function of the second kind. Since $K_i(x)$ is complex for $x<0$ we take $\mu\rho$ to be positive in order for the current to be real. We can also choose $\mu$ to be positive to eventually set $\mu=1$ which means that $\rho$ is also positive.  Notice that $x^i K_i(x)$ (no summation over $i$) is divergent at $x=0$ only for $i=0$  which means that $D \ge 4$. Notice also that near $|\vec{x}_a|$ the current goes like $|\vec{x}-\vec{x}_a|^{D-2}$ so its localized around $|\vec{x}_a|$. 
We can also integrate the current over $\eta^\mu$ to obtain
\begin{align}
	J_a(\vec{x}) = \int d\eta \, \delta^\prime(\eta^2+\mu^2) J_a(\eta,\vec{x}) =  q_a \, \mu^{D-4} \, \delta(\vec{x}-\vec{x_a}), 
\end{align}
showing that from the space-time point of view it is localized at $\vec{x}_a$. We can also compute the flux of this current up to the first neighborhood of the cotangent bundle 
and we find that the flux is constant and non vanishing
\begin{align}\label{g2}
	\Phi_a = \int d\vec{x} \, d\eta \, \delta^\prime(\eta^2+\mu^2) \, J_a(\eta,\vec{x}) = - q_a \mu^{D-4}, 
\end{align}
and it is independent of $\rho$.

Finally we can compute the interaction energy due to the disturbances of the vacuum. Inserting (\ref{g1}) in  (\ref{c4}) and integrating in $\eta^\mu$ we find 
\begin{align}\label{g3}
E =  q_1 \, q_2 \frac{\mu^{D-4} }{(D-2) \, (2\pi)^\frac{D-1}{2}} \times \hspace*{7.5cm} 
\nonumber \\
\times \left[ (D-2) \left( \frac{\mu\rho}{\sqrt{D-2}} \right)^\frac{D-1}{2} \frac{1}{|\Delta \vec{x}|^\frac{D-5}{2}}  K_\frac{D-1}{2} \left( \frac{\mu\rho}{\sqrt{D-2}} |\Delta \vec{x}| \right)
 \right. \hspace*{1.5cm} \nonumber \\ 
\left. - \left( 1+(D-2)(D-3) \right) \left(\frac{\mu\rho}{\sqrt{D-2}|\Delta \vec{x}|} \right)^\frac{D-3}{2} 
K_\frac{D-3}{2} \left( \frac{\mu\rho}{\sqrt{D-2}} |\Delta \vec{x}| \right) 
 \right], 
\end{align}
where $|\Delta\vec{x}| = |\vec{x}_2-\vec{x}_1|$.
%
%
%
At short distances $|\Delta\vec{x}| << (\mu\rho)^{-1}$ we find 
\begin{align}
	E =  -  q_1 \, q_2  \frac{\mu^{D-4} \, \Gamma\left( \frac{D-3}{2}\right)}{16(D-2) \,\pi^{\frac{D-1}{2}}} \, \frac{1}{|\Delta\vec{x}|^{D-3}},  
	\label{g8}
\end{align}
so that the force is attractive for charges $q_1$ and $q_2$ of the same sign and it is independent of $\rho$. For large distances $|\Delta\vec{x}| >> (\mu\rho)^{-1},  $
\begin{align}\label{g5}
	E =  q_1 \, q_2 \frac{\mu^{D-4}}{2 \,(2\pi)^{\frac{D-2}{2}}} \left( \frac{\mu\rho}{\sqrt{D-2}} \right)^{\frac{D-2}{2}}  \frac{e^{-\frac{\mu\rho|\Delta\vec{x}|}{\sqrt{D-2}}} }{|\Delta\vec{x}|^{\frac{D-4}{2}}},
\end{align}
the force is repulsive for charges of same sign and it depends on $\rho$. Notice that at short distances we get the standard behavior for low spin massless particles since $E$ goes like $-1/|\Delta\vec{x}|^{D-3}$. At large distances however we get the surprising behavior that $E$ approaches zero as $ {e^{-\frac{\mu\rho|\Delta\vec{x}|}{\sqrt{D-2}}} }/{|\Delta\vec{x}|^{\frac{D-4}{2}}}$. 

In Fig.1 we plot $E$ for the case $D=4$ when $q_1$ and $q_2$ have the same sign. Notice that 
for $D>4$ we get a similar behavior. There is just one point of maximum for any $D$ which is the turning point where the attractive force becomes repulsive as $|\Delta\vec{x}|$ grows. Taking the derivative of $E$ with respect to $|\Delta\vec{x}|$ we find 
\begin{align}\label{g51}
	K_\frac{D-1}{2} \left( \frac{\mu\rho}{\sqrt{D-2}} |\Delta \vec{x}| \right)  - \sqrt{D-2} \,  \mu\rho \, |\Delta \vec{x}|   \, K_\frac{D-3}{2} \left( \frac{\mu\rho}{\sqrt{D-2}} |\Delta \vec{x}| \right)  = 0,  
\end{align}
which can be used to determine the turning points of $E$. 
In general (\ref{g51}) has several real and complex roots. It is quite remarkable that there is just one positive real root. Its value depends on $D$ being $|\Delta\vec{x}|= \sqrt{2}/(\mu\rho)$ for $D=4$ and growing like $\sqrt{D}/(\mu\rho)$ for $D>>1$. We could not find a closed form for it. The roots of $E$ can also be determined. Again, there are real and complex roots but just one which is real and positive. 

\begin{figure}[htb]
\begin{center}
\includegraphics[height=3in,width=3in]{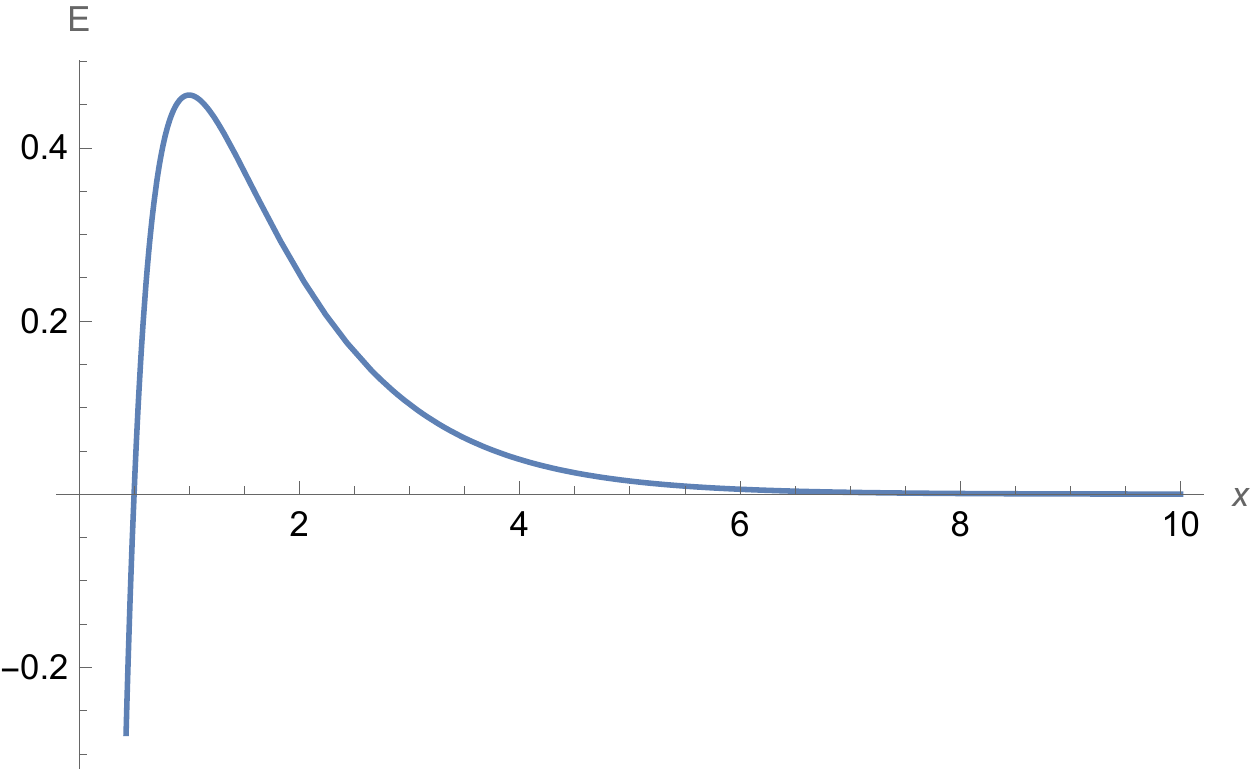}
\caption{$E \times x$ graphics for $D=4$ with $x=\frac{\mu \rho |\Delta \vec{x}|}{\sqrt{2}}$ dimensionless and charges $q_1$ and $q_2$ of the same sign. $E$ has a maximum at $x=1$ and  vanishes for finite $x$ at $x=1/2$. For other values of $D$ we have a similar behavior. }
\end{center}
\end{figure}


For $D=4$ (see Fig.1) our results are quite simple since 
\begin{align}\label{g6}
	E = \frac{q_1 \, q_2}{8\pi} \left( \sqrt{2}{\mu\rho} - \frac{1}{|\Delta\vec{x}|} \right) e^{-\frac{\mu\rho}{\sqrt{2}}|\Delta\vec{x}|}. 
\end{align}
For $|\Delta \vec{x}| < (\sqrt{2}\mu\rho)^{-1}$ the charged particles attract each other if $q_1$ and $q_2$ have the same sign while for $|\Delta \vec{x}| > (\sqrt{2}\mu\rho)^{-1}$ the charges repel each other. At $|\Delta \vec{x}| = (\sqrt{2}\mu\rho)^{-1}$ there is no force between them.

Back to the general $D$ case we find that for $\rho=0$ both terms in (\ref{g3}) are non vanishing and we get 
\begin{align}\label{g7}
	E = - q_1 \, q_2 \, \frac{\mu^{D-4} \, \Gamma(\frac{D-3}{2})}{4 \,(D-2)\, \pi^\frac{D-1}{2}} \frac{1}{|\Delta \vec{x}|^{D-3}}, 
\end{align}
so that charged particles with the same sign always attract each other. For $\rho=0$ the free continuous spin field reduces to a massless higher spin theory with helicities going from zero to infinity \cite{Schuster:2014hca,Rivelles:2014fsa}. Since the no-go theorems \cite{Weinberg:1964ew,Weinberg:1965rz,Porrati:2012rd} show that higher spin fields can not mediate long range interactions we expect that for $\rho=0$ only the lower helicities survive in (\ref{g7}). 



\section{ 
Conserved Current Degrees of Freedom }

Now we want to compute the generating functional (\ref{c2}) with the propagator (\ref{b22}) to show that the exchanged continuous spin field carries all integer helicities, each one of them appearing just once. 

Taking into account that the current $J(\eta,x)$ will be expanded around the hyperboloid $\eta^2+\mu^2=0$ let us introduce new coordinates $(|\eta|, \hat{\eta}^\mu), \,\, \eta^\mu=|\eta| \hat{\eta}^\mu$, satisfying the constraint $\hat{\eta}^2=-1$, so that $\hat{\eta}^\mu$ parametrizes points on the hyperboloid while $|\eta|$ parametrizes the hyperboloids \cite{Rivelles:2018tpt}. It must be noticed that these new coordinates must be handled carefully due to the constraint imposed on them. In the new coordinates the current $J(\eta,x)$ can be expanded as
\begin{align}
	J(\eta,x) = \sum_{n=0}^{\infty} \frac{1}{n!} \, (\mu^2 - |\eta|^2)^n \, J^T_n (\hat{\eta},x), \label{f1}  \\
	J^T_n (\hat{\eta},x) = \sum_{p=0}^\infty \frac{1}{p!} \hat{\eta}^{\mu_1} \cdots \hat{\eta}^{\mu_p} 
	J^T_{\mu_1\dots\mu_p}(x), 
	\label{f2}
\end{align}
where $J^T_{\mu_1\dots\mu_p}(x)$ is 
traceless due to the constraint $\hat{\eta}^2 =-1$. 
We can now use the $\Xi$ symmetry of (\ref{e2}) and expand its parameter as in (\ref{f1}) so that all terms in the expansion (\ref{f1}) can be removed leaving only the first two terms  
\begin{align}
	J(\eta,x) = J_0^T(\hat{\eta},x) + (\mu^2-|\eta|^2) \, J_1^T(\hat{\eta},x). 
	\label{f3}
\end{align}

The current conservation equation (\ref{e3}) can then be written in momentum space as 
\begin{align}\label{f4}
\delta(\mu^2-|\eta|^2) \left[ \left( \rho - \frac{1}{\mu} ik\cdot\partial_{\hat{\eta}} \right) {\tilde J}_0^T(\hat{\eta},k) - 2\mu \, ik\cdot\hat{\eta} \, {\tilde J}_1^T(\hat{\eta},k) \right] = 0,  	
\end{align}
showing that ${\tilde J}_0^T$ and ${\tilde J}_1^T$ are not independent. Since we are on-shell we can choose $k_+$ as the only non-vanishing component of the momentum and use (\ref{f2}) to solve (\ref{f4}) as 
\begin{align}
	{\tilde{J}^T}_{0,-\mu_1\dots\mu_n}(k_+) = \frac{\mu\rho}{ik_+} \, {\tilde{J}^T}_{0,\mu_1\dots\mu_n}(k_+), \label{f5} 
	\\
	{\tilde{J}^T}_{1,\mu_1\dots\mu_n}(k_+) = - \frac{n}{2\mu^2} \, {\tilde{J}^T}_{0,\mu_1\dots\mu_n}(k_+). \label{f6}
\end{align}
Then  $\tilde{J}^T_1$ can be expressed in terms of $\tilde{J}^T_0$ so that 
\begin{align}\label{f7}
	{\tilde{J}^T}_{0,(-)^p \, \overline{\mu}_1\dots\overline{\mu}_n}(k_+) = \left(\frac{\mu\rho}{ik_+}\right)^p \, {\tilde{J}^T}_{0,\overline{\mu}_1\dots\overline{\mu}_n}(k_+), 
\end{align}
where $\overline{\mu}=(+,i)$. This means that all {\it minus} components of $\tilde{J}^T_0$ can be removed and only the $+$ and $i$ components of $\tilde{J}^T_0$  are left. We can now consider the traceless condition of $\tilde{J}^T_0$ which reads
\begin{align}\label{f8}
	2 \tilde{J}^T_{0,+-\mu_1\dots\mu_n}(k_+) = \tilde{J}^T_{0,ii\mu_1\dots\mu_n}(k_+), 
\end{align}
and use (\ref{f7}) in it to get
\begin{align}\label{f9}
	\tilde{J}^T_{0,+\mu_1\dots\mu_n}(k_+) = \frac{ik_+}{2\mu\rho} \, \tilde{J}^T_{0, i i \mu_1\dots\mu_n}(k_+). 
\end{align}
This last relation together with (\ref{f5}) allow us to write all $+$ and $-$ components of $\tilde{J}^T_0$ in terms of those with {\it i} components only. Thus the independent components of the current are 
 $\tilde{J}^T_{0,i_1\dots i_n}(k_+)$ which are traceless in their $SO(D-2)$ indices.

Using the propagator (\ref{b22}) we can write $W[J]$ (\ref{c2}) as 
\begin{align}
	W[J] = -\frac{1}{2} \int \frac{dk}{k^2} \, I(k_+), \label{f10} 
\end{align} 
where 
\begin{align} \label{f11}
  I(k_+) = I(k)\Big|_{k^2=0}, \qquad	I(k) =  \int d\eta \, \delta^\prime(\eta^2+\mu^2) \, 
	\tilde{J}(\eta,k) \, \tilde{J}(\eta,-k). 
\end{align}
The next step is to expand $\tilde{J}(\eta,k_+)$ over $\eta$ and perform the integral in (\ref{f11}). Using the results of Appendix A we find 
\begin{align}\label{f13}
	I(k_+) 
= \frac{1}{2} \mu^{D-4} \sum_{n=0}^\infty \frac{(-1)^n}{n!} \frac{(D+2n-2)(D+2n)}
{\prod\limits_{\ell=0}^{n} (D+2\ell)}  \, {\tilde{J}}_0^{T \, \mu_1 \dots \mu_n}(k_+) \tilde{J}^T_{0 \,\mu_1\dots\mu_n}(-k_+). 
\end{align}
We can now use (\ref{f5}) and (\ref{f9}) to eliminate the $+$ and $-$ components of $\tilde{J}^T_0$ leaving only its $SO(D-2)$ indices. After some work (\ref{f13}) can be rewritten as 
\begin{align}\label{f14}
	I(k_+) = \frac{1}{2} \mu^{D-4} \sum_{n=0}^\infty \frac{(D-2)!! }{(D+2n-4)!!} \sum_{p=0}^{n} \frac{(-1)^p}{2^p (n-p)!} \sum_{q=0}^{p} \frac{1}{q!(p-q)!} {\tilde J}^{T(q)}_{i_1\dots i_{n-p}}(k_+) \, {\tilde J}^{T(p-q)}_{i_1\dots i_{n-p}}(-k_+),
\end{align}
where $q$ in ${\tilde J}^{T(q)}_{i_1\dots i_{n}}$ means $q$ traces of ${\tilde J}^T_{i_1\dots i_{n}}$ in its $SO(D-2)$ indices. It must also be noticed that ${\tilde J}^{T(q)}_{i_1\dots i_{n}}$ is not $SO(D-2)$ traceless. 
Another important point is that a term like ${{\tilde J}^{T(p)}}_{i_1\dots i_n}(k_+) {\tilde J}^{T(q)}_{i_1\dots i_n}(-k_+)$ appears just once in the summation.

If (\ref{f14}) is to describe the exchange of a continuous spin field it must be possible to rewrite it as a sum of $SO(D-2)$ traceless symmetric tensors with ranks from zero to infinity, each rank appearing just once. Let us denote this tensor as $\tilde{\cal J}^t_{i_1\dots i_n}$ where $t$ means traceless only in the $SO(D-2)$ indices. This tensor must be obtained by taking traceless linear combinations of ${\tilde J}^T_{i_1\dots i_n}$ and also of his traces ${\tilde J}^{T(p)}_{i_1\dots i_n}$
\begin{align}\label{f15}
	{\tilde{\cal J}}^t_{i_1\dots i_n} = \sum_{p=0}^\infty \alpha_p^n {\tilde J}^{T \, (p)}_{i_1\dots i_n}. 
\end{align}
Then we must compute   
\begin{align}\label{f16}
\frac{1}{2} \mu^{D-4} \sum_{n=0}^\infty   \frac{1}{n!} \frac{(D-2)!!}{(D+2n-4)!!}  {{\cal{\tilde J}}^t}_{i_1\dots i_n}(k_+) {{\cal{\tilde J}}^t}_{i_1\dots i_n}(-k_+), 
\end{align}
express it in terms of $\tilde{J}^{T(p)}_{i_1\dots i_n}$ and compare with (\ref{f14}) to find the coefficients $\alpha^n_p$. As remarked before, in (\ref{f14}) each tensor product ${\tilde J}^{T(p)}_{i_1\dots i_n}(k_+) \tilde{J}^{T(q)}_{i_1\dots i_n}(-k_+)$ appears just once but in (\ref{f16}) it may appear more than once. Comparing the first few terms of both expansions we can find the coefficients $\alpha_p^n$ for low values of $p$ and $n$ which suggests a simple pattern for the higher values of $p$ and $n$ 
\begin{align}\label{f17}
	\alpha_p^n =  \frac{(-1)^p}{(2p)!!} \frac{(D+2n-4)!!}{(D+2n+2p-4)!!}. 
\end{align}
Then (\ref{f16}) can be rewritten as 
\begin{align}\label{f18}
\frac{1}{2}\mu^{D-4} (D-2)!! \sum_{n=0}^{\infty} (D+2n-4) \, \sum_{p=0}^{[n/2]} \left(-\frac{1}{2}\right)^p \frac{(D-2p+2n-6)!!}{p! (n-2p)!} \times \nonumber \\
\times \sum_{a,b=0}^\infty \frac{(-1)^{a+b}}{(2a)!!(2b)!!} \frac{1}{(D+2n+2a-4)!!(D+2n+2b-4)!!} \tilde{J}^{T(a+p)}_{i_1\dots i_{n-2p}}(k_+) \tilde{J}^{T(b+p)}_{i_1\dots i_{n-2p}}(-k_+). 
\end{align}

It is not possible to compare (\ref{f14}) and (\ref{f18}) directly since the range of the summations are quite different. However we can compare the coefficients for the currents in both cases. So lets us consider a generic term $\tilde{J}^{T(\ell)}_{i_1\dots i_r}(k_+) \, \tilde{J}^{T(m)}_{i_1\dots i_r}(-k_+)$ with $\ell\ge m$. Its coefficient in (\ref{f14}) is 
\begin{align}\label{f19}
	\frac{(-1)^{l+m} (D-2)!!}{2^{D/2+2\ell +2m+r-2}(D/2+\ell +m+r-2)!\, \ell! \, m! \, r!}, 
\end{align}
while in (\ref{f18}) it is
\begin{align}\label{f20}
	\frac{(-1)^{\ell +m} (D-2)!!}{2^{D/2+2\ell +2m+r-2} r!} \sum_{p=0}^m \frac{(-1)^p (D/2+2p+r-2)(D/2+p+r-3)!}{p!(\ell -p)!(m-p)!(D/2+p+r+ \ell -2)!(D/2+p+r+m-2)!},  
\end{align}
so if they are the same we must have
\begin{align}\label{f21}
	\sum_{p=0}^m \frac{(-1)^p (D/2+2p+r-2)(D/2+p+r-3)!}{p!(\ell -p)!(m-p)!(D/2+p+r+ \ell -2)!(D/2+p+r+m-2)!} = \nonumber \\ 
 \frac{1}{ \ell! m! (D/2+\ell +m+r-2)!}, \qquad \ell\ge m. 
\end{align}
This is an equation which depends on $\ell, m,$ and $r$. For low numeric values of $m$ and any arbitrary  value of the others two parameters we can show that this equation holds using symbolic computation software. 
If we use only numerical values for all parameters then (\ref{f21}) was satisfied for all choices that were made. 

It is also remarkable that 
(\ref{f21}) can be rewritten in terms of the generalized hypergeometric function  ${}_3F_2(a_1,a_2,a_3;b_1,b_2;x)$ as 
\begin{align}\label{f22}
	 (D/2+r-2)! \left[  \frac{ {}_3F_2(D/2+r-2,-\ell,-m, D/2+\ell+r-1,D/2+m+r-1; -1)!}{ ({(D/2+\ell+r-2)!(D/2+m+r-2)!} )}  \right.  && \nonumber \\
	\left. - \frac{2 \,\ell \, m \,\, {}_3F_2(D/2+r-1,1-\ell,1-m,  D/2+\ell+r, D/2+m+r; -1)} {(D/2+\ell+r-1)!(D/2+m+r-1)!} \right] &=& \nonumber \\ 
	 \frac{1}{(D/2+\ell+m+r-2)!}, \qquad \ell\ge m.  
\end{align}
Again, giving numerical values for $\ell,m$ and $r$ we found that (\ref{f22}) is always satisfied.  Unfortunately we were not able to find any property of hypergeometric functions that allows us to show that  equation (\ref{f22}) holds for all values of $\ell,m$ and $r$. All these results provide a very strong evidence that the exchanged fields carries the right number of degrees of freedom for a continuous spin particle.

\section{Conclusions \label{s5}}

We have shown that like the photon and the graviton the continuous spin particle mediates long range interactions. We have found the interaction energy between two static particles having charges of the same sign which are coupled to a continuous spin field. This shows that continuous spin fields can have consistent long range interactions with matter closing a gap in our understanding of massless particles and long range physics. It would be interesting to extended theses results to the other kinds of continuous spin particles like the fermionic continuous spin particle \cite{Najafizadeh:2015uxa}, the supersymmetric spin particle \cite{Najafizadeh:2019mun} and the tachyonic spin particle \cite{Rivelles:2018tpt}. 

It also quite interesting that at small distances charges of the same sign are attractive while at large distances they repel each other. Since continuous spin fields can propagate in (A)dS spaces  \cite{Metsaev:2016lhs,Metsaev:2017ytk,Metsaev:2019opn,Metsaev:2021zdg}
 it is quite important to extend our formalism to (A)dS and search for a connection involving dark matter and dark energy.


\appendix
\section{Appendix}

Since integrals over the hyperboloid like $\int d\eta \,\,  \delta(\eta^2+\mu^2) \, \eta^{\mu_1} \cdots \eta^{\mu_n}$ are divergent they must be regulated. As discussed in detail in Appendix A of \cite{Schuster:2014hca} we can regulate these integrals by analytic continuation in $\eta$ space taking Wick rotated coordinates $[\eta]^\mu=(i\eta^0, \eta^1, \dots , \eta^{D-1})$. To keep (engineering) dimension counting right we choose the normalization $\int d^D[\eta] \, \delta(-[\eta]^2+\mu^2) = \mu^{D-2}$. In \cite{Schuster:2014hca} just the the cases with low values of $n$ were given. Here we present the results for arbitrary values of $n \ge 2$ 
\begin{align}\label{f12}
\int d[\eta] \, \delta^\prime(-[\eta]^2+\mu^2) [\eta]^{\mu_1} \cdots [\eta]^{\mu_{2n}} =  \qquad\qquad\qquad\qquad\qquad\qquad\qquad\qquad\qquad   \nonumber  \\
	\frac{1}{2} (-1)^n \mu^{D+2(n-2)} \frac{(D-2)(D-4)}{\prod\limits_{\ell=0}^{n} (D-4+2\ell)} ( g^{\mu_1\mu_2} \cdots g^{\mu_{2n-1}\mu_{2n}} + permutations ), 
\end{align}
where the permutations have weight 1.

\acknowledgments

This work was supported by FAPESP Grant 2019/21281-4.

\bibliography{mybibstrings}

\end{document}